# The Application of Data Mining to Build Classification Model for Predicting Graduate Employment

Bangsuk Jantawan[*]
Department of Tropical Agriculture and International Cooperation
National Pingtung University of Science and Technology
Pingtung, Taiwan
jantawan4@hotmail.co.th

Cheng-Fa Tsai
Department of Management Information Systems
National Pingtung University of Science and Technology
Pingtung, Taiwan
cftsai2000@yahoo.com.tw

*Abstract*—Data mining has been applied in various areas because of its ability to rapidly analyze vast amounts of data. This study is to build the Graduates Employment Model using classification task in data mining, and to compare several of data-mining approaches such as Bayesian method and the Tree method. The Bayesian method includes 5 algorithms, including AODE, BayesNet, HNB, NaviveBayes, WAODE. The Tree method includes 5 algorithms, including BFTree, NBTree, REPTree, ID3, C4.5. The experiment uses a classification task in WEKA, and we compare the results of each algorithm, where several classification models were generated. To validate the generated model, the experiments were conducted using real data collected from graduate profile at the Maejo University in Thailand. The model is intended to be used for predicting whether a graduate was employed, unemployed, or in an undetermined situation.

I. INTRODUCTION

Graduates employability remains as national issues due to the increasing number of Graduates produced by higher education institutions each year. According to the United Nations Educational Scientific and Cultural Organization report, enrollment in higher education more than doubled over the past two decades from 68 million in 1991 to 151 million in 2008. At the same time, the financial crisis that began in 2008 has resulted in increasing unemployment, as highlighted in International Labor Organization's Global Employment Trends reports. The global unemployment rate was 6.2 percent in 2010 compared to 5.6 percent in 2007. According to the 2012 report, young people continue to be the hardest hit by the job crisis with 74.8 million youth being unemployed in 2011, an increase of more than 4 million since 2007 [1].

With many economies being reported as not generating sufficient employment opportunities to absorb growth in the working-age population, a generation of young productive workers will face an uncertain future unless something is done to reverse this trend. To increase the graduates' chances of obtaining decent jobs that match their education and training, universities need to equip their students with the necessary competencies to enter the labor market and to enhance their capacities to meet specific workplace demands [1].

As Thailand, there were 320,815 graduates in 2006 with bachelors' degrees and above. This figure increased to 371,982 in 2007, about 75.02 percent of graduates in 2006 (excluding those from open universities) were employed. About 18 percent of graduates were unemployed. The proportion of employed graduates dropped to 68.65 percent in 2008 and unemployment rose to 28.98 percent [2]. Hence, preparing young people to enter the labor market has therefore become a critical responsibility for universities [1].

According to data mining is a technology used to describe knowledge discovery and to search for significant relationships such as patterns, association and changes among variables in databases [3]. There are several of data mining techniques that can be used to extract relevant and interesting knowledge from large data. Data mining has several tasks such as classification and prediction, association rule mining and clustering. Moreover, classification is one of the most useful techniques in data mining to build classification models from an input data set. The used classification techniques commonly build models that are used to predict future data trends. There are several algorithms for data classification which include decision tree and Naïve Bayes classifiers and so on [4].

Furthermore, decision tree is one of the most used techniques, due to it creates the decision tree from the data given using simple equations depending mainly on calculation of the gain ratio, which gives automatically some sort of weights to attributes used, and the researcher can implicitly recognize the most effective attributes on the predicted target. As a result of this technique, a decision tree would be built with classification rules generated from it [5], and another classification that is Naïve Bayes classifier. This classification is used to predict a target class. It depends on calculations of probabilities, namely Bayesian theorem. Because of this use, results from the classifier are more accurate and more efficiency as well as more sensitive to new data added to the dataset [5].

Therefore, the aim of this research is to predicting of graduate employment has been employed, unemployed or others within the first twelve months after graduation, the raw data received from the Planning Division Office of Maejo University in Thailand. With experiment realized through a data classification that classifies a graduate profile as employed, unemployed or others. Subsequently, the main contribution of this research is the comparison of classification accuracy between two algorithms from commonly used data mining techniques in the education domain in Waikato Environment for Knowledge Analysis (WEKA) environment.

## II. Literature Review

Several researches used data mining techniques for extracting rules and predicting certain behaviors in several areas. For example, researcher has defined the performance of a frontline employee, as his/her productivity comparing with his/her peers [6]. On the other hand, described the performance of university teachers included in his study, as the number of researches cited or published. In general, performance is usually measured by the units produced by the employee in his/her job within the given period of time [7].

Researchers like Chein and Chen [8] have worked on the improvement of employee selection, by building a model, using data mining techniques, to predict the performance of newly applicants. Depending on attributes selected from their curriculum vitae, job applications and interviews. Their performance could be predicted to be a base for decision makers to take their decisions about either employing these applicants or not. And they also used several attributes to predict the employee performance. They specified gender, age, experience, marital status, education, major subjects and school tires as potential factors that might affect the performance. Then they excluded age, gender and marital status, so that no discrimination would exist in the process of personal selection. As a result for their study, they found that employee performance is highly affected by education degree, the school tire, and the job experience.

Moreover, researchers also are identified three major requirements concerned by the employers in hiring employees, which are basic academic skills, higher order thinking skills, and personal qualities. The work is restricted in the education domain specifically analyzing the effectiveness of a subject, English for Occupational Purposes in enhancing employability skills [9], [10]. Subsequently, Kahya [11] also searched on certain factors that affect the job performance. The researcher reviewed previous studies, describing the effect of experience, salary, education, working conditions and job satisfaction on the performance. As a result of the research, it has been found that several factors affected the employee's performance. The position or grade of the employee in the company was of high positive effect on his/her performance. Working conditions and environment, on the other hand, had shown both positive and negative relationship on performance. Highly educated and qualified employees showed dissatisfaction of bad working conditions and thus affected their performance negatively. Employees of low qualifications, on the other hand, showed high performance in spite of the bad conditions. In addition, experience showed positive relationship in most cases, while education did not yield clear relationship with the performance.

More recently, in Malaysia proposes a new Malaysian Engineering Employability Skills Framework, which is constructed based on requirement by accrediting bodies and professional bodies and existing research findings in employability skills as a guideline in training package and qualification of country. Nonetheless, not surprisingly, graduates employability is rarely being studied especially within the scope of data mining, mainly due to limited and authentic data source available [12].

Employability issues have also been taken into consideration in other countries. Research by the Higher Education Academy with the Council for Industry and Higher Education in United Kingdom concluded that there are six competencies that employers observe in individual who can transform the organizations and add values in their careers [13]. The six competencies are cognitive skills or brainpower, generic competencies, personal capabilities, technical ability, business or organization awareness and practical elements. Furthermore, it covers a set of achievements comprises skills, understandings and personal attributes that make graduates more likely to gain employment and successful in their chosen occupations which benefits the graduates, the community and also the economy.

However, data mining techniques have indeed been employed in education domain, for instance in prediction and classification of student academic performance using Artificial Neural Network [14], [15] and a combination of clustering and decision tree classification techniques [14]. Experiments in [16] classifies students to predict their final grade using six common classifiers (Quadratic Bayesian classifier, 1-nearest neighbour (1-NN), k-nearest neighbor (k-NN), Parzen-window, multilayer perceptron (MLP), and Decision Tree). With regards to student performance, researchers have discovered individual student characteristics that are associated with their success according to grade point averages (GPA) by using a Microsoft Decision Trees classification technique [17]. In addition, Kumar and Chadha [18] have shown some applications of data mining in educational institution that extracts useful information from the huge data sets. Data mining through analytical tool offers user to view and use current information for decision making process such as organization of syllabus, predicting the registration of students in an educational program, predicting student performance, detecting cheating in online examination as well as identifying abnormal/erroneous values.

Accordingly, the study showed a positive relationship between affiliation motivation and job performance in Malaysia. They have tested the influence of motivation on job performance for state government employees of country. As people with higher affiliation motivation and strong interpersonal relationships with colleagues and managers tend to perform much better in their jobs [19].

Tair and El-Halees [20] used data mining to improve graduate student's performance, and overcome the problem of low grades of graduate students using association, classification, clustering and outlier detection rules. Similar to the study of Bhardwaj and Pal [21] in which a data model was used to predict student's performance with emphasis on identifying the difference of high learners and slow learners using byes classification.

Decision tree as a classification algorithm has been utilized in [22] to predict the final grade of a student in a particular course. The same algorithm has been applied in Yadav, Bharadwaj, and Pal [23] on past student performance data to generate a model to predict student performance with highlights on identifying dropouts and students who need special attention and allow teachers to provide appropriate advising or counseling. Conversely, Pandey and Pal [24] have considered the qualities the teacher must possess in order to determine how to tackle the problems arising in teaching, key points to be remembered while teaching and the amount of knowledge of the teaching process. In the course of identifying significant recommendations, John Dewey's principle of bipolar and Reybem's tri-polar education systems have been used to establish a model to evaluate the teacher ship on the basis of student feedback using data mining. Consequently, While [25] also used classification technique to build models to predict new applicant's performance, they used the same to forecast employee's talents [26], [27], [28], [29].

Another technique called fuzzy has been applied in [30] build a practical model for improving the efficiency and effectiveness of human resource management while [31] has improved and employed it to evaluate the performance of employees of commercial banks.

Generally, this paper is a preliminary attempt to use data mining concepts, particularly classification, to help supporting the human resources directors and decision makers by evaluating employees' data to study the main attributes that may affect the employees' performance. The paper applied the data mining concepts to develop a model for supporting the prediction of the employees' performance. In section 2, a complete description of the study is presented, specifying the methodology, the results, discussion of the results. Among the related work, we found that work done by [11] is most related to this research, whereby the work mines historical data of students' academic results using different classifiers (Bayes, trees, function) to rank influencing factors that contribute in predicting student academic performance.

## III. METHODOLOGY

The major objective of the proposed methodology is to build the classification model that classify a graduate profile as employed, unemployed or undetermined using data sourced from the Maejo University in Thailand for 3 academic years, which consists of 11,853 instances. To build the classifiers, we combines the Cross Industry Standard Process for Data Mining methodology [32] and Process of Knowledge Discovery in [33] in which data mining is a significant step. The iterative and sequence of steps are shown in figure 1. It consists of five steps, which include Business understanding, data understanding, data preparation, modeling, evaluation and deployment along with the data discovery processes such as data cleaning, data integration, data selection, data transformation, data mining, evaluation and presentation.

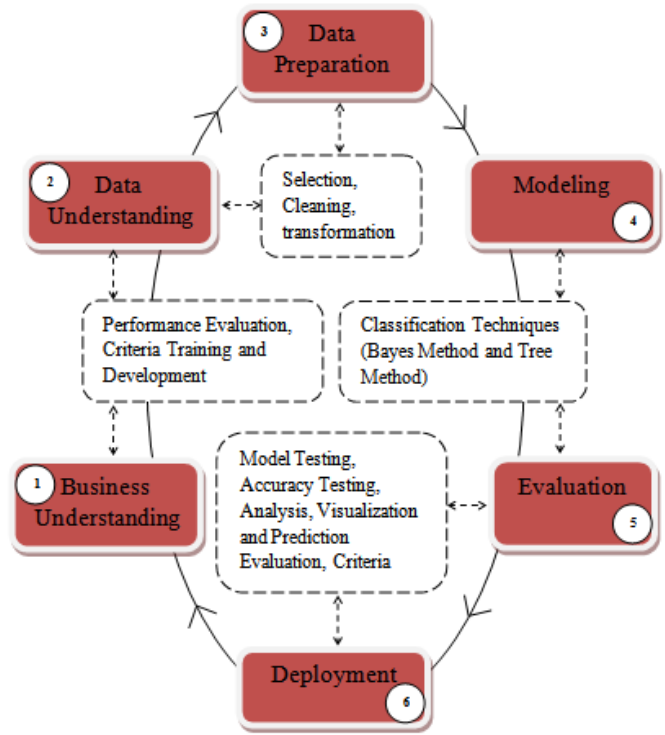

Figure 1. Framework of KDD.

### A. Business Understanding or Data Classification Preliminaries

There are two-step processes of data classification. First step is training data, which called the learning step; a model that describes a predetermined set of classes or concepts is built by analyzing a set of training database instances. Each instance is assumed to belong to a predefined class. The second step is testing data; the model is tested using a different data set that is used to estimate the classification accuracy of the model. If the accuracy of the model is considered acceptable, the model can be used to classify future data instances for which the class label is not known. Finally, the model acts as a classifier in the decision making process. There are several techniques that can be used for classification such as decision tree, Bayesian methods and so on.

Decision tree classifiers are quite popular techniques because the construction of tree does not require any domain expert knowledge or parameter setting, and is appropriate for exploratory knowledge discovery. Decision tree can produce a model with rules that are human-readable and interpretable. Decision Tree has the advantages of easy interpretation and understanding for decision makers to compare with their domain knowledge for validation and justify their decision. Some of decision tree classifiers are C4.5/C5.0/J4.8, NBTree, and others [4].

The C4.5 technique is one of the decision tree families that can produce both decision tree and rule-sets; and construct a tree for the purpose of improving prediction accuracy. The C4.5/C5.0/J48 classifier is among the most popular and powerful decision tree classifiers. C4.5 creates an initial tree using the divide-and-conquer algorithm. The full description of the algorithm can be found in any data mining or machine learning books such as Han and Kamber [33].

Furthermore, the WEKA was used for data mining. WEKA was developed at the University of Waikato in New Zealand [34]. It contains a large collection of state-of-the-art machine learning and data mining algorithms written in Java. WEKA is approved widely and is one of most complete tools in data mining. As a public data mining platform, WEKA gathers many machine learning algorithms to mine data, including data pretreatment, classification, regression, cluster class, association rule mining and visualization on new interface. WEKA has become very popular with academic and industrial researchers, and is also widely used for teaching purposes. WEKA toolkit package has its own version known as J48. J48 is an optimized implementation of C4.5.

### B. Data Understanding

In order to classify a graduate profile as employed, unemployed or undetermined using data sourced from the Maejo University in Thailand for 1 academic years, the total are 5,361 instances in 2011. Table 1 shows the complete attributes for Graduate profile data source.

### C. Data-Preprocessing

The raw data received from the Planning Division Office, Maejo University at Chiang Mai Province in Northern Thailand, which required the Data Pre-processing to prepare the dataset for the classification task. First, the data source has been transferred to Excel sheets and replaced them with the mean values of the attribute. Then, cleaning data involves eliminating data with missing values in critical attributes, correcting inconsistent data, identifying outliers, as well as removing duplicate data. For example, some attributes like GPA, have been entered in continuous values. Data source from the total of 5,361 instances in the raw data, the data cleaning process ended up 3,530 instances that are ready to be mined. These files are prepared and converted to (.csv) format to be compatible with the WEKA data mining is used in building the model.

### D. Modeling

The section of modeling and experiments, the classification models have been built to predict the employment status (employed, unemployed, others) for graduate profiles. Using the decision tree technique, in this technique, the gain ratio measure is used to indicate the weight of effective of each attribute on the tested class, and accordingly the ordering of tree nodes is specified. The results are discussed in the following sections.

TABLE I. THE ATTRIBUTES OF THE GRADUATES EMPLOYMENT DATA AFTER THE PRE-PROCESSING

| No. | Attributes | Values | Descriptions |
|---|---|---|---|
| 1 | Prefix | {Male, Female, Dr., Associate Prof. Dr,..} | Prefix of graduate |
| 2 | Gender | { Male, Female} | Gender of the graduate (Male, Female) |
| 3 | Province | {Bangkok, Suratthani…} | Province of graduate |
| 4 | Degree | {Bachelor, Ph.D., Master} | Degree of graduate |
| 5 | Educational background | {B.Sc, B.L.A, M.A., M.B.A.} | Graduate background |
| 6 | Faculty | {Science, Agricultural Production, Economics,…} | Faculty of graduate |
| 7 | program | {Computer science, Information technology,} | Program of graduate |
| 8 | GPA | {Interval value} | GPA for current qualification |
| 9 | WorkProvince | {Bangkok, Suratthani…} | Province of Student's work |
| 10 | Status | {Employed, UnemployedandNotStudy,Study, …} | Work status of graduate |
| 11 | Talent | {Computer, Art, Food physical, …} | Talent of graduate |
| 12 | Position | {Chef, Trad, boss,...} | Position of graduate |
| 13 | Satisfaction | { Pleased, Lack_of_consistence,Other,..} | Satisfaction of graduate with work |
| 14 | PeriodTimeFindwork | {FourToSix, OneToThree, SevenToNine,..} | Time of find work |
| 15 | WorkDirectGraduate | {Direct,NotDirect,NoIdentify} | Matching of Graduate education with graduate work |
| 16 | ApplyKnowlageWithWork | {Moderate, NoIdentify, Much,…} | Knowledge of graduate can apply with work |
| 17 | ResonNotWork | {Soldier, Business, NotFindWork…} | The Reason that don't have work of graduate |
| 18 | ProblemOfWork | {Lack_Of_support, NoProblem…} | The problem of work |
| 19 | RequirementsOfStudy | {NoNeed, Need} | Requirements of graduate to continue to study |
| 20 | LevelOfStudyRequired | {Master, Graduate_Diploma, NoIdentify,…} | Level Required to study of Graduate |
| 21 | InstitutionNeed | {Private, Aboard, Government…} | Institution Requirement to study of graduate |

The classification model is performed in two steps, which include training and testing. Once the classifier is constructed, testing dataset is used to estimate the predictive accuracy of the classifier. Then the WEKA have 4 types of testing option, which are using the training set, supplied test set, cross validation and percentage split. If we use training set as the test option, the test data will be sourced from the same training data, hence this will decrease reliable estimate of the true error rate. In the part of Supplied test set permit us to set the test data which been prepared separately from the training data. Cross-validation is suitable for limited dataset whereby the number of fold can be determined by user. 10-fold cross validation is widely used to get the best estimate of error. It has been proven by extensive test on numerous datasets with different learning techniques.

IV. RESULTS AND DISCUSSION

Ten classification techniques Method have been applied the dataset to build the classification model. The techniques are: The decision tree with 5 versions, BFTree, NBTree, REPTree, ID3, C4.5 (J4.8 in WEKA), and Naïve Bayes classifier with 5 version such as the Averaged One-Dependence Estimators (AODE), BayesNet, HNB, NaviveBayes, the Weightily Averaged One-Dependence Estimators (WAODE).

In table 2 shows the Classification accuracy using various algorithms under Tree method in WEKA. In addition, the table provides comparative results for the kappa statistics, mean absolute error, root mean squared error, relative absolute error, and root relative squared error from the total of 1,059 testing instances. Subsequently, result of the J48 algorithm achieved the highest accuracy percentage as compared to other algorithms. The second accuracy is REPTree algorithm, BFTree, NBTree, ID3 Respectivel.

Furthermore, figure 2 also shows an example of tree structures of J48 algorithms. Graf adds nodes to the decision trees to increase predictive accuracy. Accordingly, table 3 shows the classification accuracies for various algorithms under Bayes method. The table provides comparative results for the kappa statistics mean absolute error, root mean squared error, relative absolute error, and root relative squared error from the total of 1,059 testing instances. Also, table 3 presents the WAODE algorithm achieved the highest accuracy percentage as compared to other algorithms. Despite treating each tree augmented naive Bayes equally, have extended the AODE by assigning weight for each tree augmented naive Bayes differently as the facts that each attributes do not play the same role in classification.

In addition, a performance comparison of the Bayesian and Tree methods shows that the WAODE algorithm achieved the highest accuracy of 99.77% using the graduate data set. The second highest accuracy was achieved using a tree method, the J48 algorithm, with an accuracy of 98.31%. Using the Bayes method, the AODE algorithm, was third, with a prediction accuracy of 98.30%. We found that both classification approaches were complementary because the Bayesian methods provide a better view of association or dependencies among attributes, whereas the results of the tree method are easier to interpret.

TABLE II. THE CLASSIFICATION ACCURACY USING VARIOUS ALGORITHMS UNDER TREE METHOD IN WEKA

| Algorithm | Accuracy (%) | Error Rate (%) | Kappa Statistics | Mean Absolute Error | Root Mean Squared Error | Relative Absolute Error (%) | Root Relative Squared Error (%) |
|---|---|---|---|---|---|---|---|
| ID3 | 90.20 | 9.3229 | 0.9408 | 0.0126 | 0.112 | 6.37 | 35.56 |
| J48 | 98.31 | 1.69 | 0.9586 | 0.0166 | 0.0912 | 7.886 | 28.0945 |
| BFTree | 98.24 | 1.76 | 0.9572 | 0.0165 | 0.0928 | 7.8327 | 28.5893 |
| NBTree | 98.07 | 1.93 | 0.953 | 0.01 | 0.0942 | 4.7667 | 29.0166 |
| REPTree | 98.27 | 1.73 | 0.9578 | 0.0169 | 0.0919 | 8.0104 | 28.3152 |

TABLE III. THE CLASSIFICATION ACCURACY USING VARIOUS ALGORITHMS UNDER BAYES METHOD IN WEKA

| Algorithm | Accuracy (%) | Error Rate (%) | Kappa Statistics | Mean Absolute Error | Root Mean Squared Error | Relative Absolute Error (%) | Root Relative Squared Error (%) |
|---|---|---|---|---|---|---|---|
| AODE | 98.30 | 1.70 | 0.9586 | 0.0096 | 0.0909 | 4.5744 | 28.0238 |
| BayesNet | 98.02 | 1.98 | 0.9525 | 0.0124 | 0.0911 | 5.8937 | 28.0643 |
| HNB | 97.25 | 2.75 | 0.9331 | 0.0158 | 0.1097 | 0.1097 | 33.8206 |
| NaviveBayes | 97.96 | 2.04 | 0.9509 | 0.0118 | 0.0937 | 5.607 | 28.8735 |
| WAODE | 99.77 | 0.23 | 0.9946 | 0.0028 | 0.0324 | 1.3306 | 9.9742 |

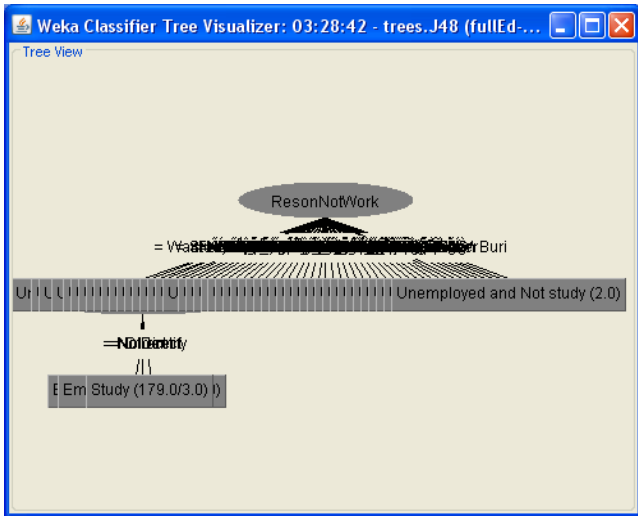

Figure 2. The tree structure for J48 algorithms.

Figure 3 shows the mapping of the root mean squared error values resulting from the classification experiment. This knowledge can be used to gain insights into the employment trend of graduates from local institutions of higher learning. A radial display of the root mean squared error across all algorithms under both Bayesian and tree-based methods reveals the accuracy of these approaches. A smaller mean squared error results in a better forecast. Based on this figure, Bayesian methods produced a better forecast than the corresponding tree methods.

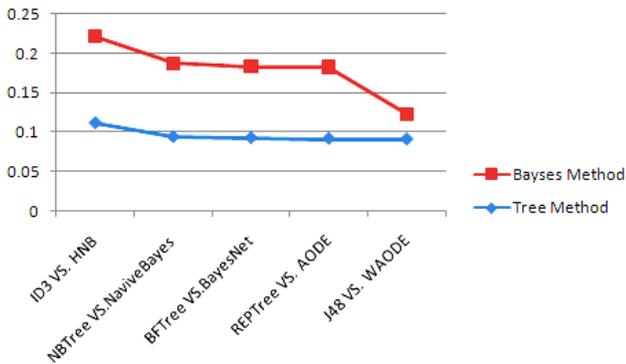

Figure 3. Mapping of the root mean squared error values of Bayesian and Tree methods.

## V. CONCLUSIONS AND FUTURE WORK

As graduates remains increase number of graduates produced by higher education institutions each year, graduates are facing more competition to ensure their employment in the job market. The purpose of the study is to assist higher-education institutions in equipping their graduates with sufficient skills to enter the job market. This study attempts to identify the attributes that influence graduate employment based on actual data obtained from the graduates themselves 12 months after graduation.

This study attempts to predict whether a graduate has been employed, remains unemployed, or is in an undetermined situation after graduation. We performed this prediction based on a series of classification experiments using various algorithms under Bayesian and decision methods to classify a graduate profile as employed, unemployed, or other. Results show that the WAODE algorithm, a variant of the Bayse algorithm, achieved the highest accuracy of 99.77%. The average accuracy of other Tree algorithms was 98.31%.

In future research, we hope to expand the data set from the tracer study to include more attributes and to annotate the attributes with information such as the correlation factor between current and previous employers. We are also looking at integrating data sets from different sources of data, such as graduate profiles from the alumni organizations of various educational institutions. We plan to introduce clustering in the preprocessing phase to cluster the attributes before attribute ranking. Finally, we may adopt other data-mining techniques, such as anomaly detection or classification-based association, to gain more knowledge of the graduate employability in Thailand. We also plan to use a data set from the National Pingtung University of Science and Technology in Taiwan and compare the results with the data set from Thailand.

ACKNOWLEDGMENT

B. Jantawan would like to express thanks Dr. Cheng-Fa Tsai, professor of the Management Information Systems Department, the Department of Tropical Agriculture and International Cooperation, National Pingtung University of Science and Technology in Taiwan for supporting the outstanding scholarship, and highly appreciates to Mr. Nara Phongphanich and the Planning Division Office, Maejo University in Thailand for giving the information.

AUTHORS PROFILE

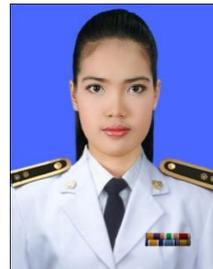

**Bangsuk Jantawan** is a Ph.D. student at the Department of Management Information System, National Pingtung University of Science and Technology (NPUST) in Taiwan, where her research is the Application of Data Mining to Build Classification Model for Predicting Graduate Employment. She obtained her MSc degree in Education Technology from the King Mongkut's University of Technology Thonburi, and a BSc degree in Computer Engineering form Dhurakij Pundit University in Bangkok, Thailand. Jantawan's prior research involved the Information System Development for Educational Quality Administration of Technical Colleges in Southern Regional Office of the Vocational Commission. Her present research interests include the data mining, education system development, decision making, and machine learning.

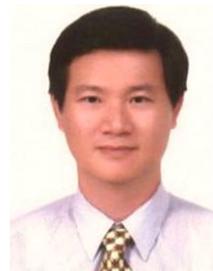

**Cheng-Fa Tsai** is full professor of the Management Information Systems Department at National Pingtung University of Science and Technology (NPUST), Pingtung, Taiwan. His research interests are in the areas of data mining and knowledge management, database systems, mobile communication and intelligent systems, with emphasis on efficient data analysis and rapid prototyping. He has published over 160 well-known journal papers and conference papers and several books in the above fields. He holds, or has applied for, nine U.S. patents and thirty ROC patents in his research areas.